# Magneto-optical studies of the uniform critical state in bulk MgB$_2$.

A.A. Polyanskii, A. Gurevich, J. Jiang, D.C. Larbalestier,

*Applied Superconductivity Center, University of Wisconsin, Madison WI 53706*

S.L. Bud'ko, D.K. Finnemore, G. Lapertot[1], P.C. Canfield

*Ames Laboratory, U.S. Department of Energy and Department of Physics and Astronomy*
*Iowa State University, Ames, Iowa 50011.*

*Abstract*

We present a detailed magneto-optical investigation of the magnetic flux penetration in polycrystalline MgB$_2$ slabs made by direct reaction of B and Mg. Our results unambiguously indicate a uniform, Bean critical state magnetization behavior with almost no electromagnetic granularity. From the measured magnetic flux profiles we were able to extract the temperature dependence of the critical current density J$_c$(T). The J$_c$(T) value reaches $1.8 \times 10^5$ A/cm$^2$ at 10K and 0.12T, in good agreement with global magnetization measurements.

## Introduction

The discovery that MgB$_2$ has a critical temperature T$_c \approx$ 40K [1] has stimulated considerable interest in the properties of this "intermediate-T$_c$", weakly anisotropic polycrystalline material. Basic thermodynamic parameters of MgB$_2$ have been measured [2-6], and high critical current densities J$_c \sim 10^5 - 10^6$ A/cm$^2$ have been reported from resistive and magnetization measurements for bulk reacted and bulk sintered samples [3-5], and for wires [7,8]. Even higher J$_c$ values $\sim 10^5 - 10^7$ A/cm$^2$ and enhanced irreversibility fields have been determined for thin films [9-12]. It has also been deduced from many magnetization and some transport measurements that MgB$_2$ does not appear to exhibit either weak-link electromagnetic behavior at grain boundaries [13] or the fast flux creep [14] which limit the performance of high-T$_c$ superconductors. These features of MgB$_2$ are potentially very valuable for large-scale applications.

However, the sintering or reaction of MgB$_2$ into fully dense bodies is still comparatively rare. Direct reaction of Mg and B sometimes seems to give better connected samples [2,3,6] than sintering of pre-reacted MgB$_2$ [13], leading to variable behavior on local scales. In most measurements so far, high J$_c$ values have been deduced from global measurements of either transport critical current or total magnetic moment, from which J$_c$ is then calculated using the Bean model. These methods do not give much information about the actual current distribution, for which only *local* techniques, such as Hall probe, or magneto-optical (MO) imaging are suitable, at least down to scales (~5-10 µm) at which they still retain resolution [15]. MO studies on early sintered [13] and directly reacted samples [16,17] have shown considerable departures from a uniform critical state. However, other directly reacted [15] and hot isostatically pressed [18] MgB$_2$ samples exhibit more uniform MO images, consistent with uniform current flow that can be analyzed in terms of the Bean model.

In this paper we report on a detailed MO investigation of a directly reacted MgB$_2$ sample which shows a rather uniform critical state, very unlike the percolative current flow revealed by MO imaging of high-T$_c$ superconductors [19-21]. Because the grain size is 3-5 µm is below the resolution limit of the MO technique, it is possible to further support the studies already cited [2-4,6,8,13] that conclude, by more indirect means, that grain boundaries are not weak links which cause electromagnetic granularity on more macroscopic scales. The MO technique also enables us to extract the temperature dependence of the local critical current density $J_c(T,B)$ and the magnetic field of first flux penetration $H_f(T)$, which can then be compared to previous reports of the lower critical field $H_{c1}$ derived from global magnetization measurements. A brief report on our work was made earlier [15].

---
[1] On leave from Commissariat a l'Energie Atomique, DRFMC-SPSMS,
38054 Grenoble, France

## Experimental Details

We studied bulk, isotopically pure $Mg^{10}B_2$ samples synthesized by direct reaction of Mg and B powders as described in Refs. [2,3]. The material, whose global magnetization and transport properties were investigated in Refs. [2,3], has a sharp resistive transition ($\Delta T < 0.5$ K) at $T_c = 40.2$ K, as measured by a SQUID. The MO imaging was performed at magnetic fields $0 < H < 160$ mT and temperatures $10 < T < 42$ K on a sample, which was cut into a 3.3×3×0.6 mm rectangular slab without problems due to spalling of the material. The MO experimental setup is described in Ref. [20]. We used a set of garnet MO indicator films of relatively large thickness of 5 µm made from $(Bi,Lu)_3(Fe,Ga)_5O_{12}$ with in-plane magnetization and different saturation fields 90 mT and 200 mT. The indicator film was placed directly onto the sample surface. For each set of MO measurements, the indicator films were calibrated by measuring the relation between the MO light intensity and the applied magnetic field at 45K. We also performed scanning electron microscopy (SEM) using specimens prepared by high-precision mechanical polishing with diamond films and low-speed polishing wheels

## Results and discussion

Fig. 1 shows typical MO images of the magnetic flux distributions in the sample when the broad face (3×3.3 mm) was oriented perpendicular to the applied field. Here Fig. 1a shows magnetic flux penetration in the zero-field cooled (ZFC) sample at 20, 35 and 40K, while Fig. 1b shows MO images of the trapped flux after an applied field of 160 mT was turned off. The MO images clearly exhibit regular field patterns consistent with closed, uniformly spaced current streamlines that rotate by 90° at each corner, producing the roof-top distribution of magnetization predicted by the Bean model. The distributions of measured perpendicular field component $B_z(x,y)$ at the sample surface appear rather uniform with no visible macroscopic defects such as cracks, surface notches, etc., and no indication of electromagnetic granularity [19-21] on the scale above the 5-10 µm MO spatial resolution. This resolution is below the characteristic grain size ~ 2 µm observed by SEM, but is sufficient to reveal the sample inhomogeneities on the scales ~ 30-50 µm.

For more quantitative studies, we have measured magnetic flux profiles $B_z(x)$ across the slab sample in a magnetic field H applied parallel to its broad face. For this orientation, demagnetization effects are strongly reduced, which greatly simplifies the analysis of MO images. Fig. 2a shows a sequence of MO images taken in this parallel geometry, while Fig. 2b shows the corresponding flux density profiles $B_z(x)$ (b) across the ZFC slab at 10, 35, 37, and 39K after applying a parallel field of 80 mT. The flux profiles $B_z(x)$ were taken in the region between two vertical lines in Fig. 2a, and then averaged in the y-direction over the length ≈ 100 µm to reduce the effect of noise in the MO indicator film and some blurring of the MO response caused by the ~20% porosity in the sample (see Fig. 5). At low temperatures the sample exhibits incomplete flux penetration (the slight bump in $B_z(x)$ near x = 0 is due to magnetic domains in the MO indicator film). As expected, the slope $dB_z/dx$ decreases as T increases, and the magnetic flux reaches the center of the sample only at higher temperatures at which the applied field H exceeds the field of full flux penetration $H_p = J_c(T)d/2$, where d = 0.6 mm is the sample thickness.

Shown in Fig. 3 are MO images (a) and the corresponding averaged flux profiles (b) for a sequence of field-increasing and field-decreasing steps after H was first increased to 100 mT and then decreased to zero. After field reversal, the magnetization currents change direction, producing triangular distributions of trapped flux. It can be seen that the magnitude of the flux gradients is rather constant, whatever the applied field, a result that implies little sensitivity to field over the range 0-100 mT, exactly the range of field for which polycrystalline, untextured high temperature superconductors exhibit strong weak link effects

[19,20]. By contrast, Figs. 1-3 all indicate the classical uniform, almost field-independent behavior of a Bean critical state.

The use of the parallel slab geometry enabled us to extract the local critical current density $J_c(T)$ from the measured slope of $H_z(x)$ at the sample surface, using the Maxwell equation, $J_c = 2dH_z/dx$. Here the factor 2 accounts for the fact that the MO technique measures the field $H_z(x)$ at the surface of a long (semi-infinite) slab [21].

Fig. 4a shows the temperature dependence of $J_c(T)$ at 0.12 Tesla obtained from the measured profiles $H_z(x)$ for $12 < T < 39$ K. The sample was first zero-field cooled and then a parallel field H = 0.12 Tesla was applied. The resulting flux profiles $H_z(x)$ were averaged over the length ≈ 100 µm in the y-direction, as described above. The slope $dH_z/dx$ was taken over a length ≈ 80 µm, in the region about 50 µm away from the sample edge, where $H_z(x)$ profiles were rather straight. The so-obtained $J_c$ data presented in Fig. 4a can be well described by the formula

$$J_c(T) = J_0[1 - (T/T_c)^2]^a, \qquad (1)$$

where $J_0 = 0.2$ MA/cm$^2$, and a = 0.85. These data can be compared with the results of global magnetization measurements on a different piece of the same material, for which the field dependence $J_c(H)$ at different T was obtained from SQUID measurements [3] (Fig. 5b). The $J_c(T)$ dependence in Fig. 4a derived from the MO analysis is qualitatively consistent with the magnetization $J_c(T)$ data shown in Fig. 4b, except that the MO $J_c$ values are some 50% higher. This discrepancy is not surprising given natural material variability and the effect of the sample shape, which complicate quantitative interpretation of the global magnetization measurements of Ref. [3]. By contrast, the use of the well-defined slab geometry of the MO experiment avoids uncertain geometrical factors of the Bean model, enabling us to measure local $J_c$ values.

Although the MO images appear rather uniform on the scale of the slab thickness, one can see roughening of the magnetic flux penetration front and variations of the local slope $dH_z/dx$ on scales ~ 30 - 50 µm. Analysis of SEM images for this sample indicates that the inhomogeneities of MO images correlate with microstructural inhomogeneities. For instance, Fig. 5a and b show SEM images of our sample for different magnification levels. Fig 5a reveals fully reacted polycrystalline $MgB_2$ regions of size 30-50 µm separated by porous regions. The size of the fully reacted $MgB_2$ regions is of the order of the scale of the variations of the local MO intensity. From the direct weighting, we estimated the density of the sample as 2.04 g/cm$^3$, which is about 77% of the theoretical density, 2.63 g/cm$^3$ [22]. From the higher magnification SEM image in Fig. 5b, we inferred the $MgB_2$ grain size to be of order 2-5 µm.

Using the MO technique, we have also measured the minimum external field, which causes flux trapping at the sample surface. The experiment was carried out by cooling the sample in zero field, then applying an incrementally increasing magnetic field, the reducing it to zero to observe when flux was first trapped. The minimum field $H_i$ which produces a detectable surface trapped flux was defined as $H_{c1}(T)$, although $H_i$ is always higher than the real $H_{c1}(T)$ because of the effect of flux pinning and the surface barrier. Nevertheless, it is instructive to analyze this measurement of $H_i(T)$, because it places a lower limit on the London penetration depth, $\lambda(T)$. The results are shown in Fig. 6, from which $\lambda(T)$ can be extracted using the relation $H_{c1} = (\phi_0/4\pi\lambda^2)(\ln\kappa + 0.5)$, where $\phi_0$ is the flux quantum, $\kappa = \lambda/\xi$, and the coherence length $\xi$ can be obtained from the measurements of the upper critical field $H_{c2} = \phi_0/2\pi\xi^2$ [3,11]. At T = 0, we obtain $H_i \approx 28$ mT, which is rather close to the value of $H_{c1}$ derived from global magnetization measurements on a single crystal [22]. Hence, we obtain $\lambda(0) \approx 142$ nm, if we take $\kappa \approx 30$ and $\xi(0) = 5$ nm, based on the results of Ref. [3]. Our data are compared to the results from global measurements on polycrystals of $H_{c1}(T)$ extracted by other groups from the reversible component of the magnetization [3,23-25]. This value of $\lambda(0)$ is qualitatively consistent with the results of other groups [3, 23-25], and is also rather close to the value $\lambda(0) \approx 140$ nm [3] deduced from measurements of the reversible magnetization. Notice that our MO setup detects surface trapped flux up to 41.5K, the effect, which requires further studies.

**Summary**

In conclusion, our MO results prove that a rather uniform critical state can be developed in polycrystalline $MgB_2$. No magnetic granularity characteristic of high-$T_c$ superconductors was observed. The MO images do show roughening of the magnetic flux penetration front on a scale consistent with porosity in our $MgB_2$ sample. The MO technique enabled us to extract the temperature dependence of the local $J_c(T)$ values, which are in qualitative agreement with the results of global magnetization measurements.

**Acknowledgments**

The work at the University of Wisconsin has been supported by AFOSR, DOE and the NSF though the MRSEC on Nanostructured Materials. Ames Laboratory is operated for the U. S. Department of Energy by Iowa State University under Contract No. W-7405-Eng.-82. Work at Ames Laboratory was supported by the director of Energy Research, Office of Basic Energy Sciences.

**Captions**

Fig. 1. MO images of the magnetic flux distribution in an almost square (3 x 3.3 mm) $MgB_2$ slab in a perpendicular applied field: (a) flux penetration after zero field cooling in the noted fields to the noted temperatures; (b) distribution of trapped magnetic flux after the applied field was first ramped to 160 mT and then turned off.

Fig. 2. MO images (a) and corresponding magnetic flux profiles (b) after zero field cooling to 10K (1), 35K (2), 37K (3), 39K (4) in a field of H=80 mT applied parallel to the long axis of the slab. The vertical lines show the region over which the flux profiles were averaged. The two horizontal lines mark the sample width of 0.6 mm.

Fig. 3. MO images (a) and the corresponding flux density profiles for the $MgB_2$ slab in parallel fields at 37K. The field was first increased from 0 to 0.1T (curves 1-3) and then reduced towards zero (curves 4 and 5). The vertical lines show the region where the flux profiles were taken, and the horizontal lines show the sample width of 0.6 mm.

Fig. 4a. Temperature dependence of $J_c(T)$ extracted from the slopes of the measured flux profiles B(x) after a parallel field of 0.12 T was applied after zero field cooling. The MO data are shown as solid squares while the solid line was calculated from Eq. (1). The inset shows an MO image of the roof-top pattern of trapped magnetic flux near the sample edge after a parallel field H = 0.12T at 38 K was applied.

Fig. 4b. $J_c$ data obtained from global magnetization measurements of Ref.[3] with the use of the Bean model. Our MO data are shows as a dashed line.

Fig. 5. SEM images of the $MgB_2$ sample on different spatial scales.

Fig. 6. The field of the onset of MO flux penetration (unfilled circles) in comparison with $H_{c1}$ data obtained from magnetization measurements [23-25].

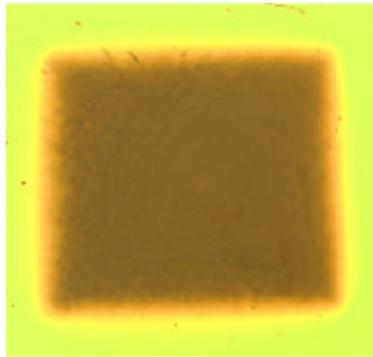

T = 20 K  H = 160 mT

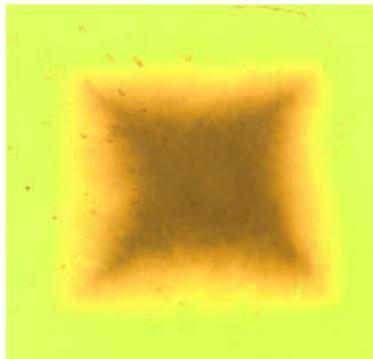

T = 35 K  H = 160 mT

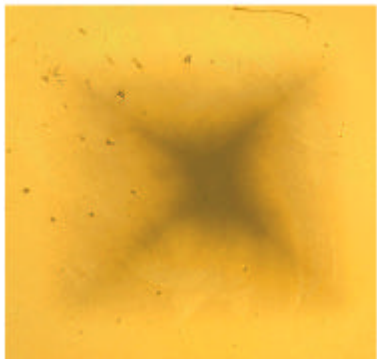

1 mm

T = 40 K  H = 4 mT

Fig. 1a

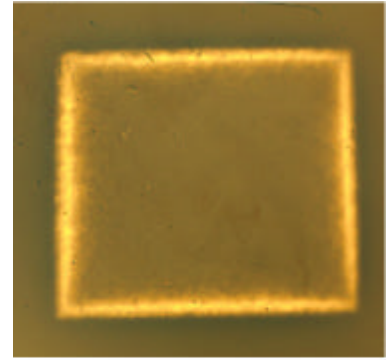

T = 20 K

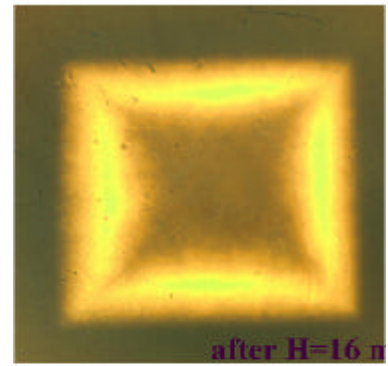

after H=16 m
T = 35 K

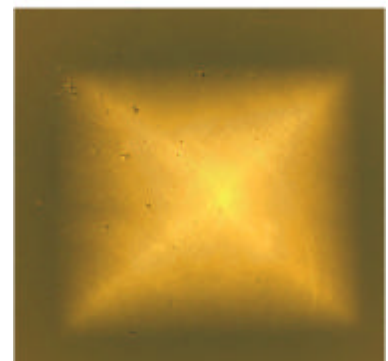

1 mm
T = 40 K

Fig. 1b

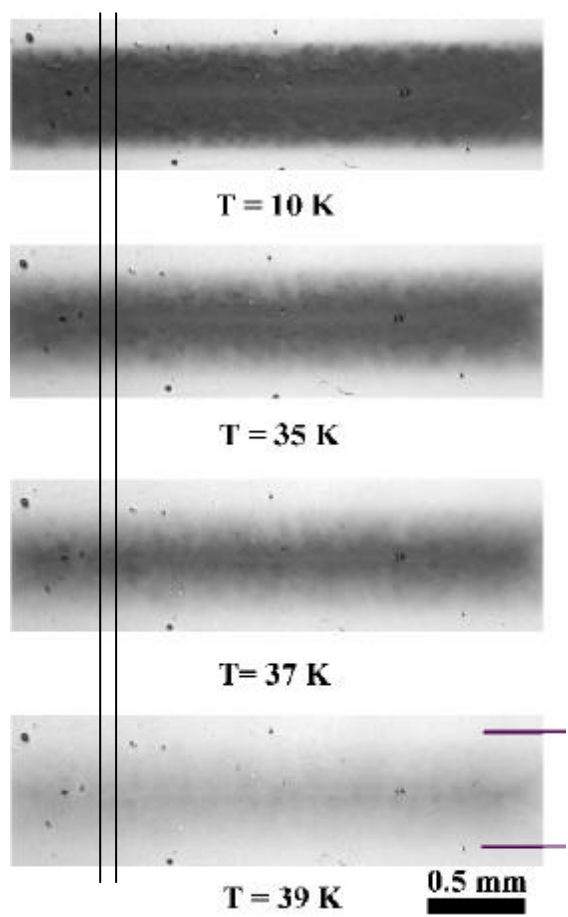

Fig. 2a

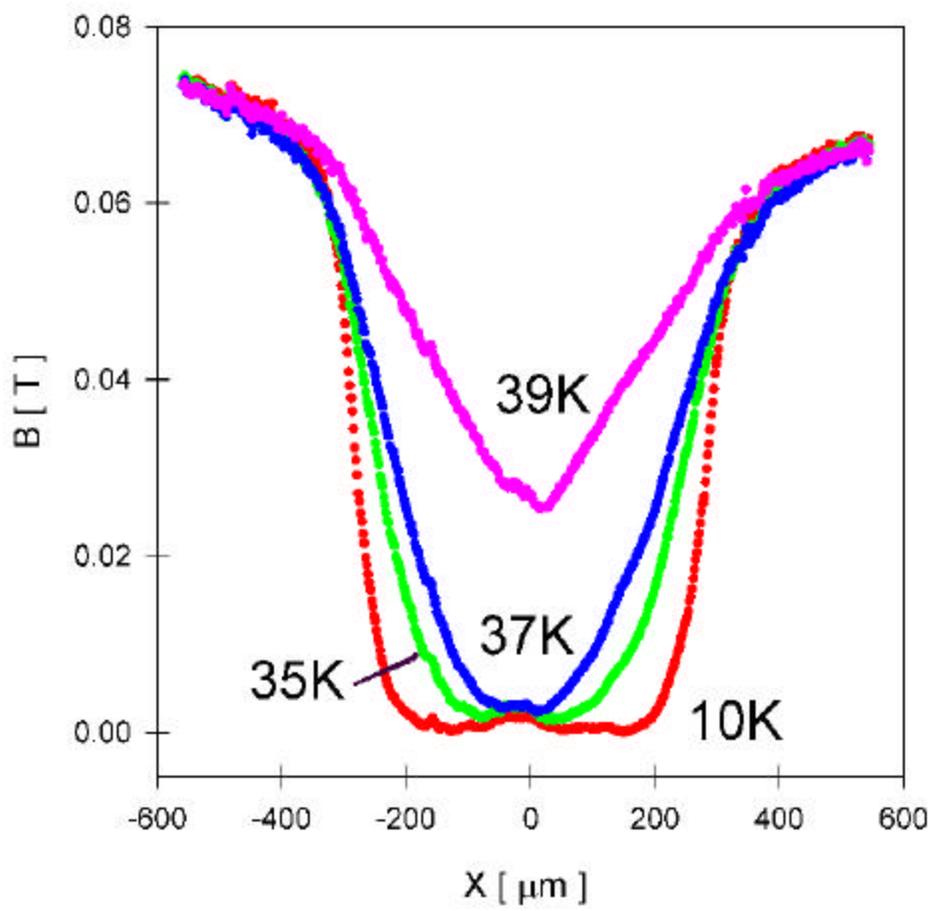

Fig. 2b

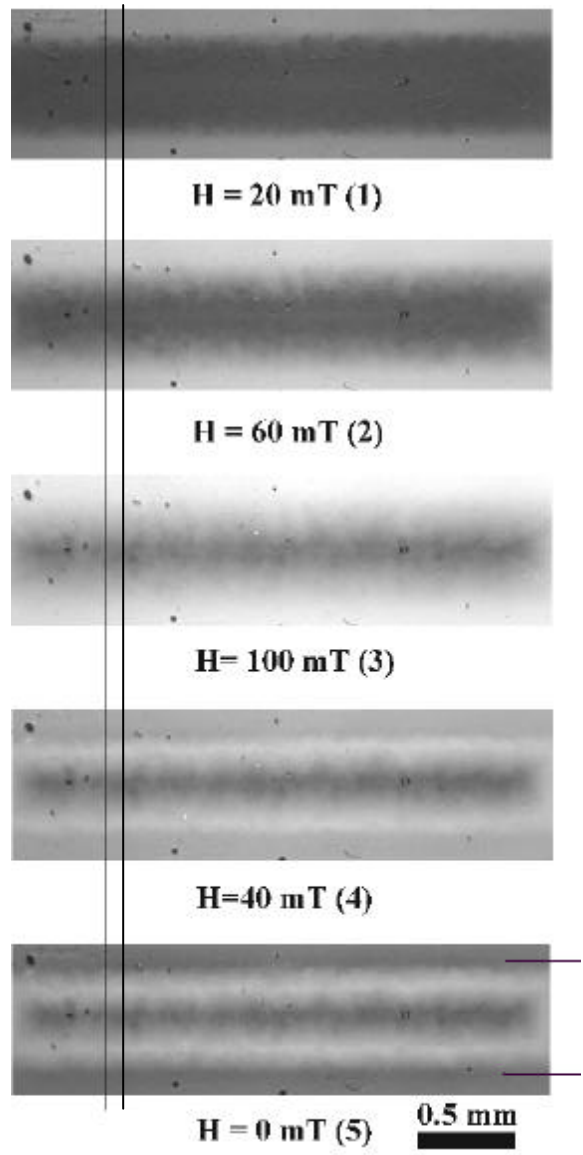

Fig. 3a

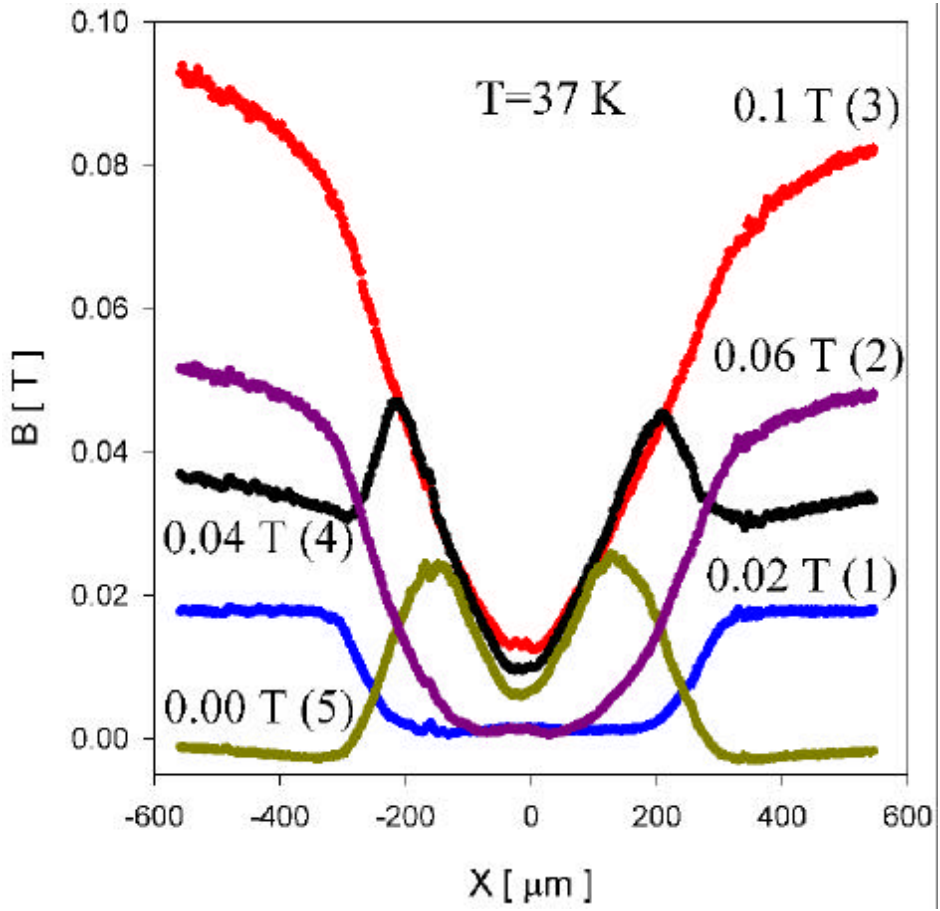

Fig. 3b

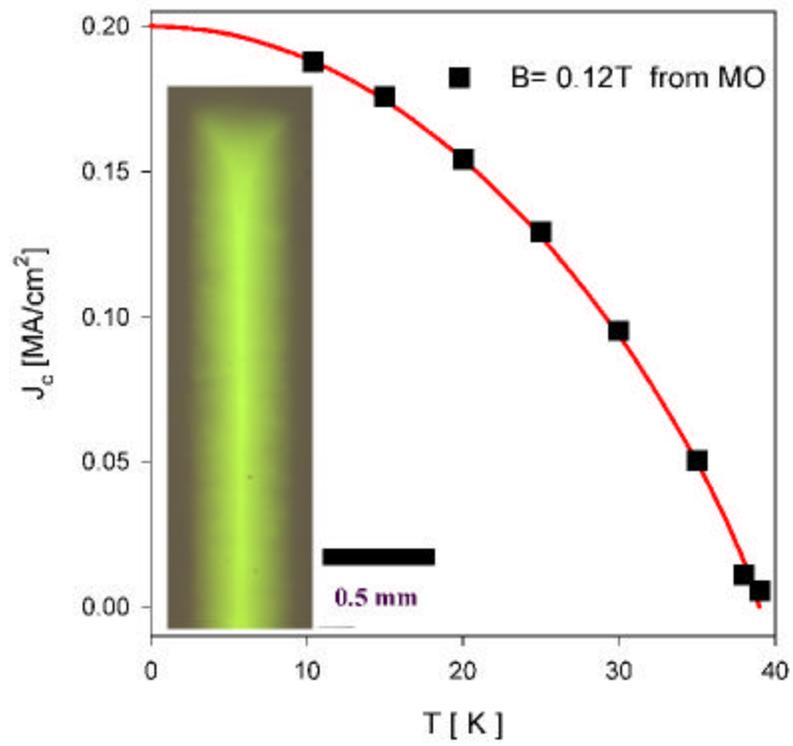

Fig.4a

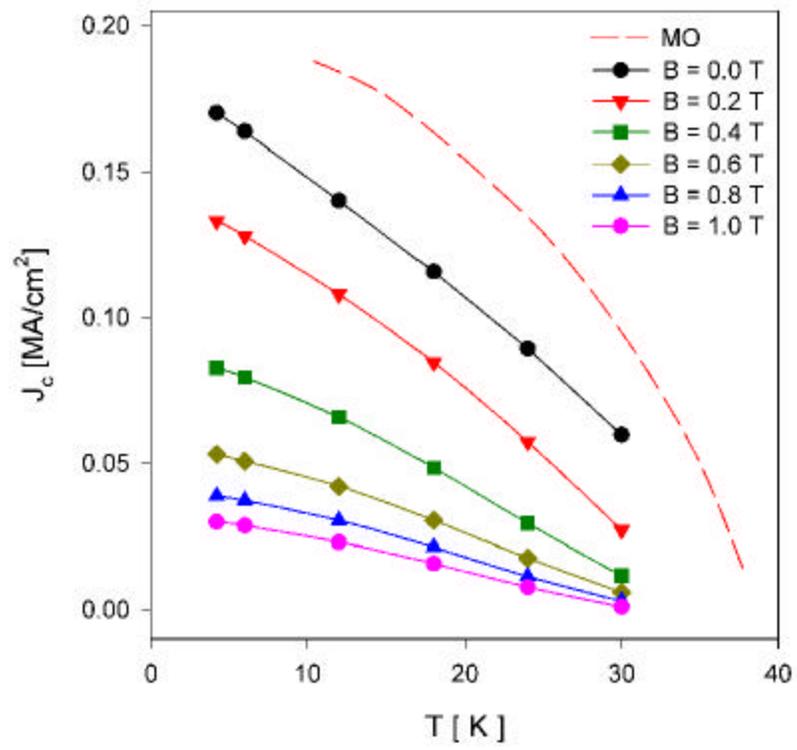

Fig.4b

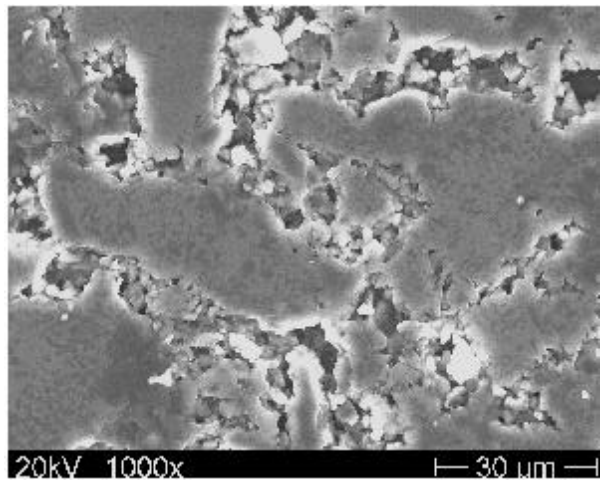

Fig. 5

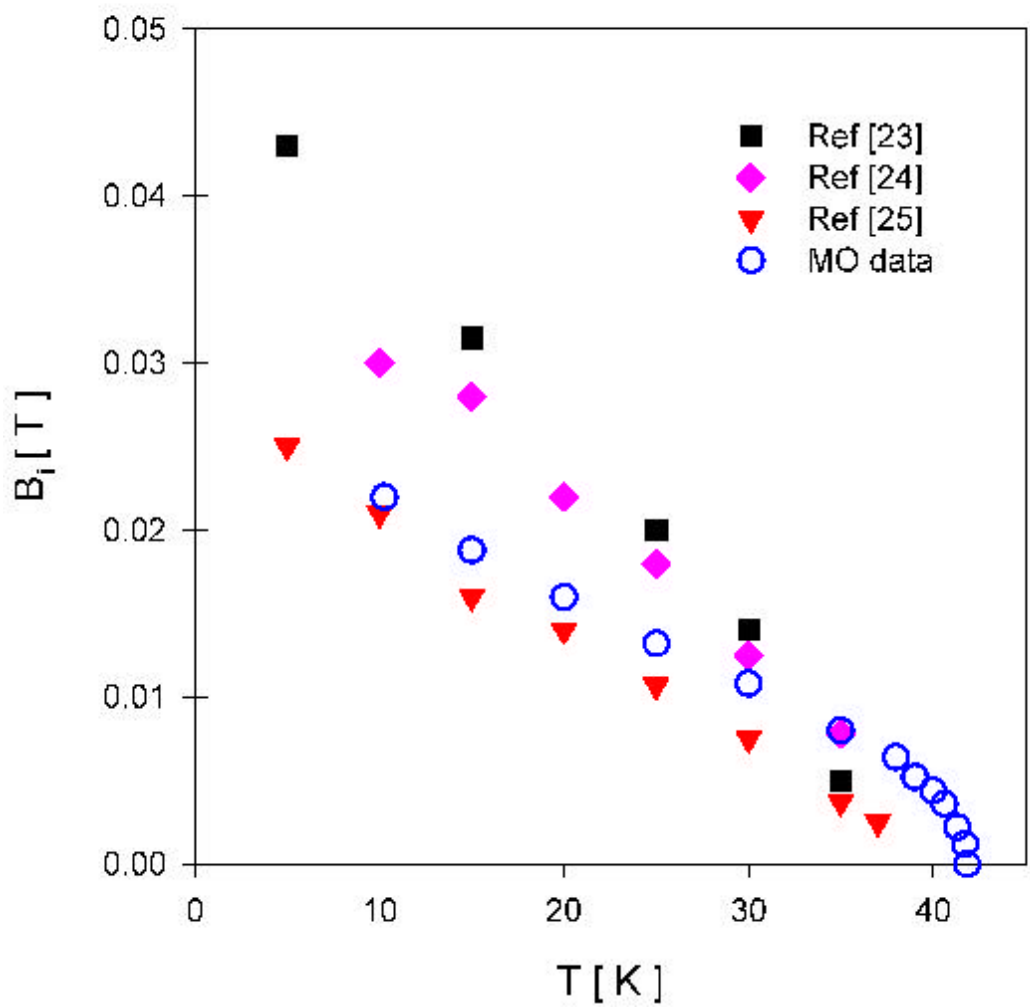

Fig. 6